\documentclass[runningheads]{llncs}
\usepackage{graphicx}
\usepackage{algorithm}
\usepackage{algpseudocode}
\usepackage{amsmath}
\usepackage{amssymb}
\usepackage{booktabs} 
\usepackage{tabularx} 
\usepackage{lipsum} 
\usepackage{adjustbox}
\usepackage{subcaption}
\usepackage{multirow}
\usepackage{xcolor}
\usepackage{breqn,xspace}
\usepackage{bm}
\usepackage{lipsum,multicol}
\usepackage{makecell}
\usepackage{amsmath}
\usepackage{url}
\usepackage{cite}
\usepackage{comment}
\begin{document}
\title{Hurry: Dynamic Collaborative Framework For Low-orbit Mega-Constellation Data Downloading}
\titlerunning{Hurry: Dynamic Collaborative Framework for LMCN Data Downloading}
%
\author{Handong Luo\and
Wenhao Liu\and
Qi Zhang \and
Ziheng Yang \and
Quanwei Lin \and
Wenjun Zhu\ \and
Kun Qiu \and
Zhe Chen \and
Yue Gao 
}
\authorrunning{H. Luo et al.}
%

\institute{Fudan University, Shanghai, China\\
\email{\{hdluo23,liuwh23,qizhang23,zhyang22,qwlin22\}@m.fudan.edu.cn, \{wenjun,gao.yue,qkun,zhechen\}@fudan.edu.cn}
}
%
\maketitle              

\begin{abstract}
     Low-orbit mega-constellation network, which utilize thousands of satellites to provide a variety of network services and collect a wide range of space information, is a rapidly growing field. Each satellite collects TB-level data daily, including delay-sensitive data used for crucial tasks, such as military surveillance, natural disaster monitoring, and weather forecasting. According to NASA's statement, these data need to be downloaded to the ground for processing within $3\sim5$ hours. To reduce the time required for satellite data downloads, the state-of-the-art solution known as CoDld, which is only available for small constellations, uses an iterative method for cooperative downloads via inter-satellite links. However, in LMCN, the time required to download the same amount of data using CoDld will exponentially increase compared to downloading the same amount of data in a small constellation. We have identified and analyzed the reasons for this degradation phenomenon and propose a new satellite data download framework, named Hurry. By modeling and mapping satellite topology changes and data transmission to Time-Expanded Graphs, we implement our algorithm within the Hurry framework to avoid degradation effects. In the fixed data volume download evaluation, Hurry achieves 100\% completion of the download task while the CoDld only reached 44\% of download progress. In continuous data generation evaluation, the Hurry flow algorithm improves throughput from 11\% to 66\% compared to the CoDld in different scenarios.
\end{abstract}

\section{INTRODUCTION}
Low-orbit Mega-Constellation Network (LMCN) is an emerging and rapidly developing field. In recent years, constellation projects like Starlink~\cite{noauthor_starlink_nodate}, OneWeb~\cite{noauthor_eutelsat_nodate}, and Starshield~\cite{noauthor_spacex_nodate} have been proposed and are gradually being deployed. These large low-Earth-orbit (LEO) constellations utilize tens of thousands of satellites orbiting the Earth to provide services such as Internet communication, Earth observation, and space data collection~\cite{yuan2024satsense}.

Each LEO satellite generates approximately $1$TB of data daily~\cite{tao2023transmitting,ma_remote_2015}, containing information for services like military surveillance~\cite{korody2004satellite}, natural disaster monitoring~\cite{visser2004real}, and weather forecasting~\cite{geer2018all}. Due to the limited processing capabilities of satellites~\cite{theis_spacecraft_1983,lin2023fedsn,yuan2023graph}, this data needs to be downloaded to the ground for processing within 3-5 hours~\cite{earth_science_data_systems_lance_2021} according to NASA's statement. How to rapidly download data from high-speed moving satellites to the ground through fixed ground stations is a challenging problem.

Current research largely focuses on using ground-satellite links (GSL)~\cite{hou2019satellite} for data transmission~\cite{chen2020satellite,castaing2014scheduling,tao2023transmitting,maillard2016adaptable}. However, communication between satellites and ground stations only lasts for $5\sim15$ minutes at a time, occurring $6\sim8$ times daily\cite{cakaj_practical_2009}. Utilizing inter-satellite links (ISL)~\cite{arora2017review} enables satellites to transfer data among themselves, accelerating data downloading. Research utilizing ISL has shown that the CoDld algorithm, which employs iterative methods and bipartite graph matching, can fully utilize GSL bandwidth in small constellations (typically ranging from 10 to 66 satellites) like Iridium~\cite{fossa1998overview} and Globalstar~\cite{dietrich1998globalstar}. However, in LMCN where the number of satellites ranges from 500 to 12,000~\cite{curzi2020large}, the CoDld algorithm suffers from severe degradation, with download speeds decreasing as the download progresses. The time require to download the last 25\% of the data is 326 times longer than that for the first 25\%. 

In this paper, we analyze the cause of this degradation, terming it the Proximal Station Bias Degradation (PSBD), an algorithmic degradation phenomenon where data on satellites requires more hops to reach the ground station as the downloading progresses. This is mainly due to the fact that data closer to ground stations is prioritized for downloading, while data farther away from the ground stations is suspended during its download process. To address this degradation and accelerate satellite data downloading, we propose a flow planning algorithm. This algorithm models and maps the predicted satellite dynamic topology and data generation into Time-Expanded Graphs~\cite{kohler2002time} to calculate transmission schemes. We also establish the \textsl{Hurry} framework to dynamically adjust the transmission scheme based on its actual execution.

We integrate the Hurry framework as a plugin into the Plotinus~\cite{Gao2024Plotinus}, a satellite digital twin system, to validate our algorithm's performance. Our algorithm can achieve 100\% completion of a single download task while the CoDld algorithm can only reach 44\% download progress. In experiments with continuous data generation, our algorithm can improve throughput by $11\%\sim66\%$ compared to the CoDld in different scenarios.

This paper contributes by:
\begin{enumerate}
    \item Identifying and thoroughly analyzing the PSBD phenomenon in existing satellite data downloading algorithms like CoDld, especially when applied to LMCN.
    \item Introducing \textsl{Hurry}, a dynamic satellite downloading framework designed to optimize satellite data download time, implementing a transmission scheme generation algorithm based on Time-Expanded Graphs~\cite{kohler2002time} within this framework.
    \item Validating the performance of the Hurry framework and the flow planning algorithm using the Plotinus, a satellite digital twin system.
\end{enumerate}

The paper is organized as follows: Section 2 describes the background and the PSBD phenomenon. Section 3 provides an overview of the Hurry framework. Section 4 gives the problem formulation, details, and the specific algorithm design. Section 5 describes the evaluation, and we conclude in Section 6.

\section{BACKGROUND}

\subsection{Constellation, Links, and Download}
In this paper, we focus on the Walker Delta satellite constellation. 
In this satellite constellation, inter-satellite links (ISL) and ground-satellite links (GSL) form the foundation of the satellite network. ISL enables direct communication between satellites, while GSL is the direct communication link between satellites and ground stations. Under the +Grid ISL structure~\cite{bhattacherjee2019network}, each satellite can establish connections with four adjacent satellites: two for connecting with adjacent satellites in the same orbital plane, and another two for connecting with satellites that have the same numerical designation in neighboring orbits.

Currently, in satellite download algorithms that only utilize GSL, the state-of-the-art method is Umbra~\cite{tao2023transmitting}, which employs network flow-based graph: Time-Expanded Graphs to manage the data volume transmitted to ground stations. The algorithm proposed in this paper is innovated by this study. The exploration of using ISL to accelerate satellite data downloading is initially put forward by the CoDld algorithm. Zhang~\cite{zhang2019energy} subsequently introduce an energy-efficient variant of CoDld~\cite{jia2017collaborative}, and Wu~\cite{wu_optimal_2019} extend the model to encompass multiple ground stations. To the best of our knowledge, all existing researches on satellite downloading are based on small constellations comprising merely a few dozen satellites (typically ranging from 10 to 200 satellites).

\begin{figure}[htbp]
  \centering
  \begin{subfigure}[b]{0.48\textwidth} 
    \includegraphics[width=\linewidth]{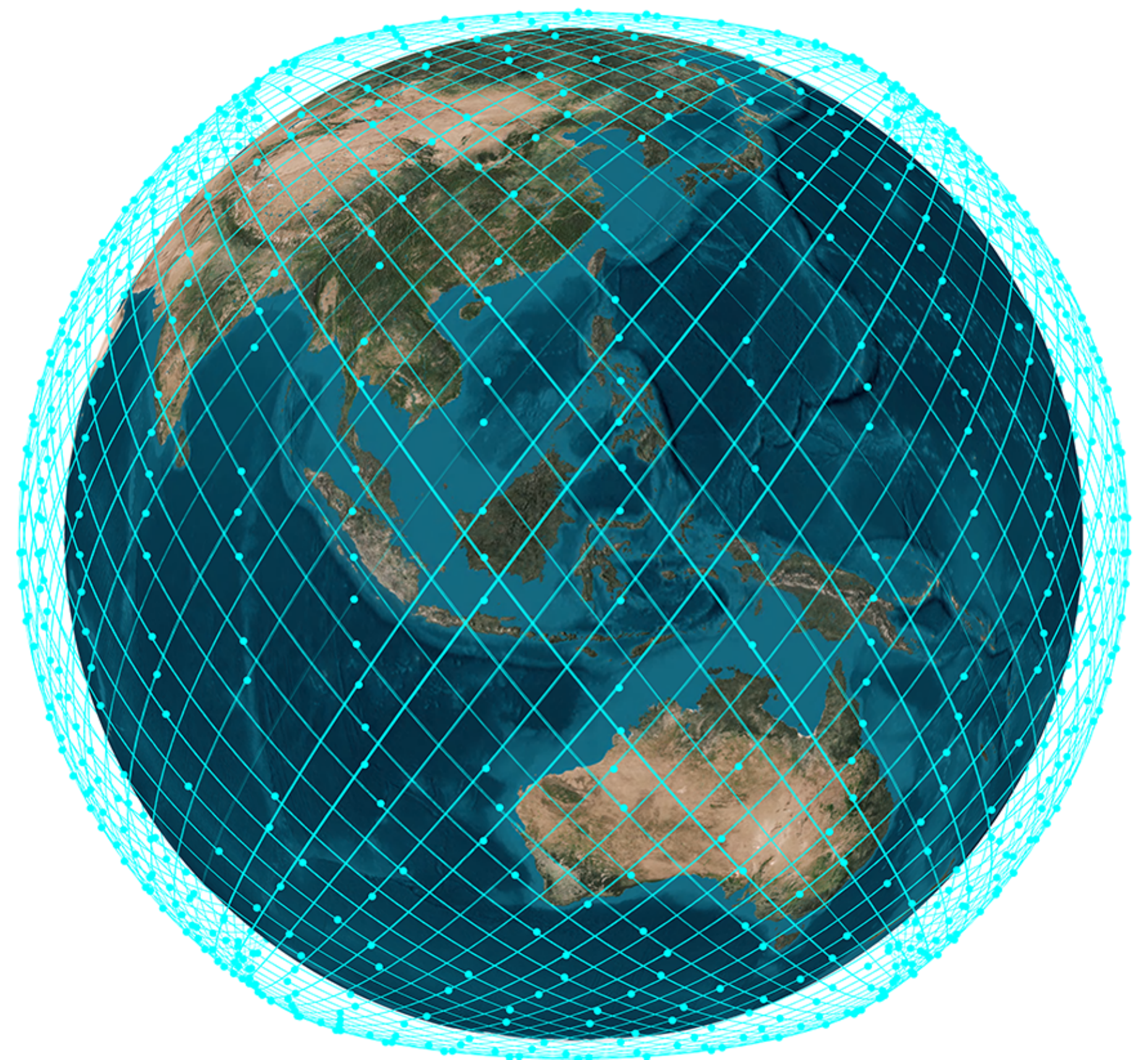} 
    \caption{Starlink \textit{shell1} constellation}
    \label{fig:shell1}
  \end{subfigure}
  \hfill 
  \begin{subfigure}[b]{0.45\textwidth} 
    \includegraphics[width=\linewidth]{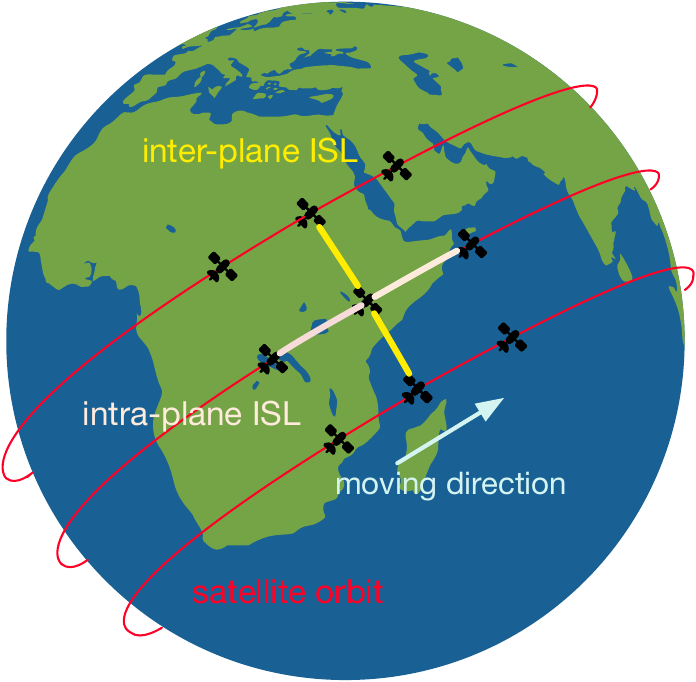} 
    \caption{+Grid ISL structure}
    \label{fig:isl}
  \end{subfigure}
  \caption{Walker Delta satellite constellation and +Grid ISL structure }
\end{figure}


\subsection{CoDld Algorithm}
The CoDld algorithm employs an iterative method to construct a download schedule for the satellite network. For $i$-th satellite \(S_i\) in the network, the algorithm maintains two key variables: remaining download time and remaining contact time with the ground station. The remaining download time indicates how long the time required by \(S_i\) to fully download its data, while the remaining contact time signifies how much ground station contact time \(S_i\) can still be allocated for communication with the ground station. During each iteration, satellites waiting to download data and satellites with spare contact time are paired by a maximum bipartite graph matching algorithm. Upon successful matching, data will be transmitted to the satellite that has available contact time through inter-satellite links, thereby being relayed to the ground. This iterative process continues until all satellites have completed their data downloads or the contact time with the ground station is fully utilized. A notable characteristic of this algorithm is that if a satellite has data yet to be downloaded, it will not accept any data from other satellites until its data download is completed.


\subsection{Proximal Station Bias Degradation}
PSBD is an algorithmic degradation phenomenon in which data from satellites requires an increasing number of hops to reach the ground station during the download process. This phenomenon is caused by the imbalanced distribution of data across the satellite network during download, where data closer to ground stations is given priority for downlink, while data farther from the ground stations remains suspended during its download process.


\begin{figure}[h]
  \centering
  \begin{subfigure}[b]{0.55\textwidth} 
    \includegraphics[width=\linewidth]{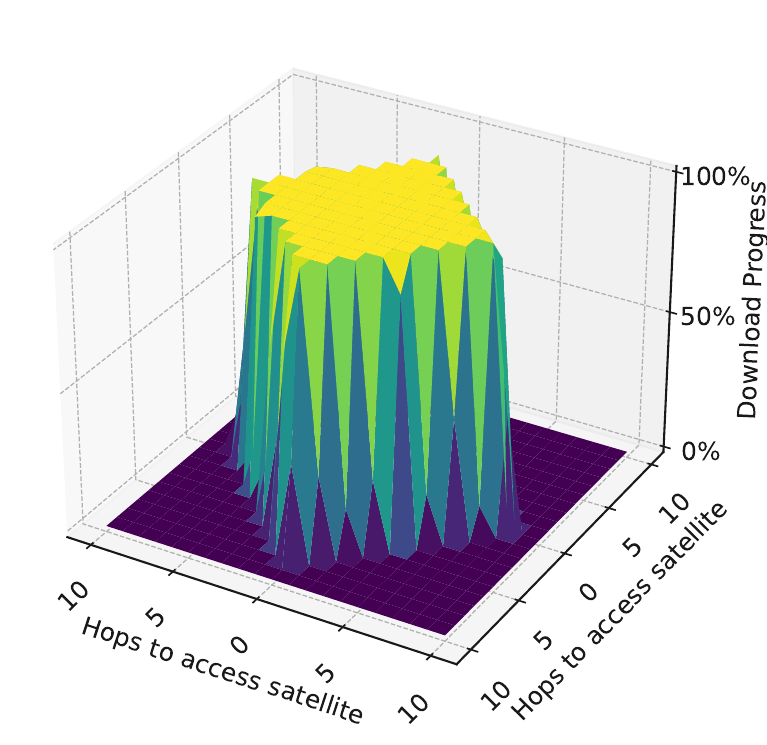}
    \caption{\textbf{Uneven distribution of data download progress across satellite constellation during PSBD events}. The x-axis and y-axis coordinates represent the number of hops needed for the satellite to reach the access satellite within the current orbital plane and between orbital planes, respectively. The z-axis coordinate represents the current satellite's data download progress.}
    \label{fig:distribution}
  \end{subfigure}
  \hfill 
  \begin{subfigure}[b]{0.43\textwidth} 
  \includegraphics[width=\linewidth]{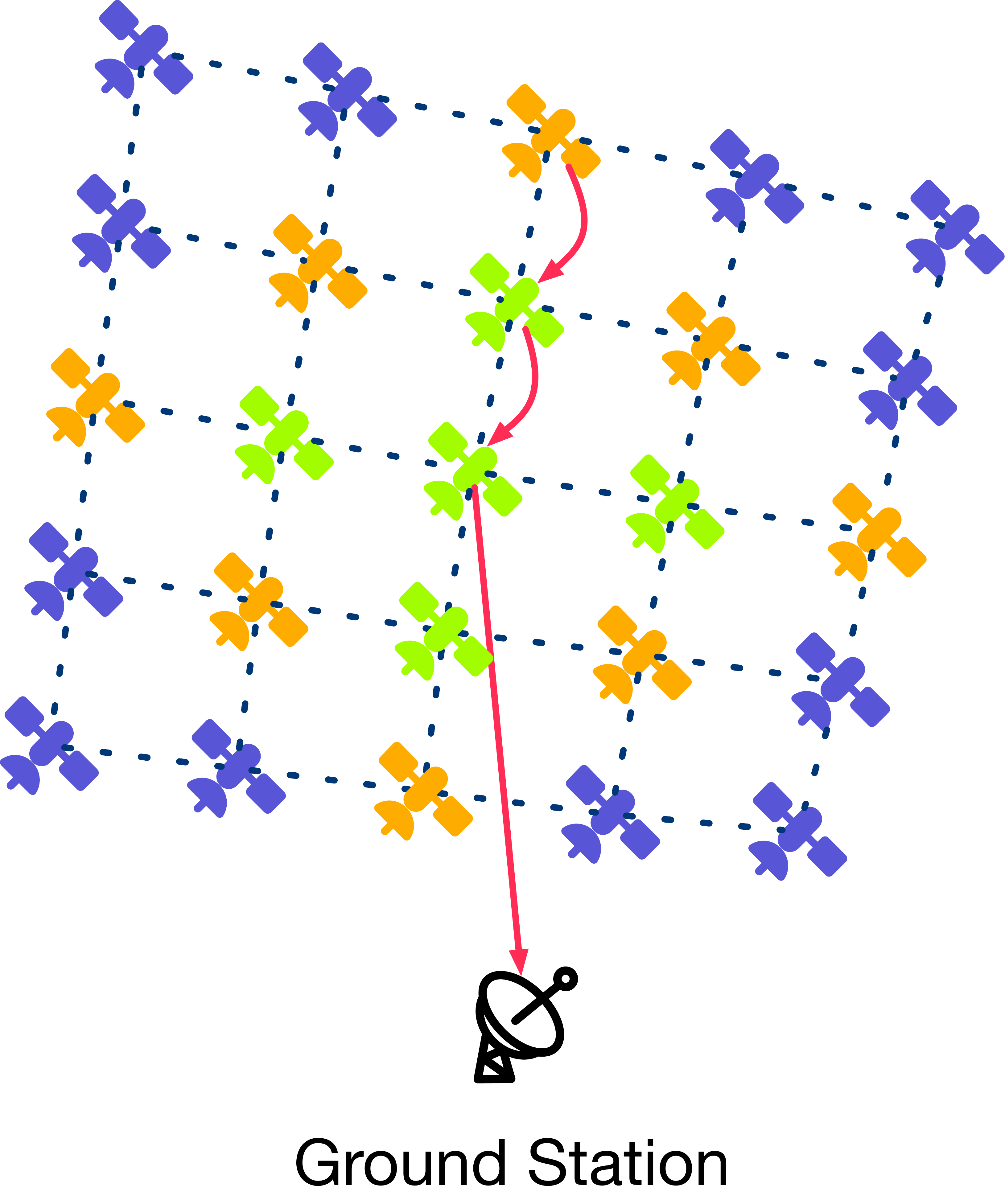}
  \caption{\textbf{Transmission process of satellite data download via inter-satellite Links}. Green satellites represent those that have completed data downloads, orange satellites represent those currently downloading data, and blue satellites represent those that have not yet started downloading data.}
  \label{fig:ideal}
\end{subfigure}
\caption[short]{The PSBD phenomenon and its consequences}
\end{figure}

Fig.~\ref{fig:distribution} shows the distribution of data download progress in the satellite network when the PSBD phenomenon occurs. Satellites closer to the ground station in Manhattan were given priority to download all satellite data, while those farther away did not begin downloading, creating an area where no data exists on the satellites. All data needed for download were located on satellites far from the ground station. Fig.~\ref{fig:ideal} shows the process of satellites distant from the ground station transmitting data to the ground station via inter-satellite links. Since data offloading only takes place on satellites one hop away, the data must survive multiple time-consuming inter-satellite offloading before it can be transmitted to the ground. The number of hops for data offloading will continue to increase as the area where no data exists on the satellites expands, further slowing down the download speed.

As we have mentioned before, existing solutions such as CoDld cannot prevent the degradation effect, which may cause a download suspension. Thus, a new algorithm that can solve PSBD needs to be proposed to increase the downloading speed in LMCN.

\section{OVERVIEW OF HURRY}

\begin{figure}[h]
    \centering
    \includegraphics[width=0.8\linewidth]{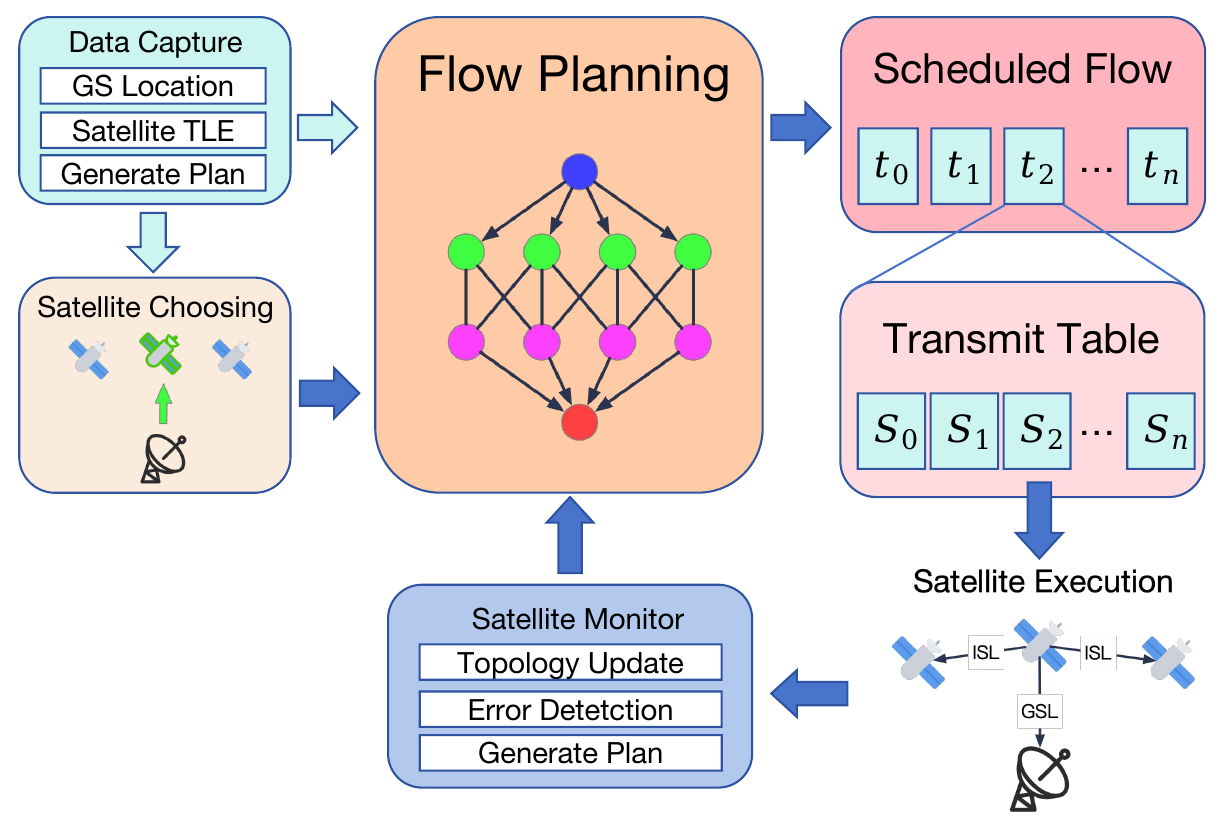}
    \caption{Overall architecture design for Hurry framework}
    \vspace{-0.4cm}
    \label{fig:arch}
\end{figure}

In this section, we will provide an overview of the Hurry framework. The overall design of Hurry, as shown in Fig.~\ref{fig:arch}, is composed of four submodules: data capture, satellite selection, flow planning, and satellite monitoring.

\textbf{Data Capture module} captures data from the satellite constellation as the initial state of our system. It collects three key pieces of data: location of the ground station, the Two Line Element (TLE) orbit descriptors of the satellite, and download plan of each satellite. The first two types of data can be easily obtained from ITU software or public websites like Space-Track or CelesTrak. TLE descriptors should be updated periodically to ensure accurate satellite orbit positioning.

\textbf{Satellite Selection module} uses its emulator to emulate the satellite position in each time slot. It calculates the distance between the satellite and ground station to determine the visible satellites for each ground station at a given time. The ground station then selects satellites in its visible satellite list to establish a data connection. After the connection is established, the satellite sends the data generated by itself or from other satellites to the ground station.

\textbf{Flow Planning module} takes the data from the Data Capture model and the Satellite Selection model as input and generates a transmission plan. It determines the direction and volume of data transmission from satellites to ground stations or between satellites.

\textbf{Satellite Monitoring module} monitors the satellite's queuing status and the unpredictable topology change of the satellite constellation. If the queue state of a satellite does not meet the predefined threshold, the monitor module sends a signal to the flow planning module to adjust the plan. This module also tracks the topology change of the satellite constellation owing to the high mobility of LEO satellite constellations.

\section{FLOW PLANNING}

\subsection{Problem Formulation}

\begin{table}[h]
  \caption{Terms of Definition}
  \centering
  \begin{tabular}{ll}
  \hline
  Notation & Definition \\
  \hline
  \( t \) & The \(t\)-th timeslot \\
  \( n \) & The total number of satellite \\
  \(m \) & The total number of ground station \\
  \( S_i \) &  Satellite \(i\) \\
  \( S_i^{t}\) & Node for Satellite \(i\) at time slot \(t\) \\
  \( G_i \) & Ground station  \(i\) \\
  \( G_i^{t} \) & Node for Ground station  \(i\) at time slot \(t\)  \\
  \( D \) & Transmit plan matrix \\
  \( D_{S_i,G_j}^{t} \) & Data amount transferred from \( S_i \) to \( G_j \) at time \(t\) \\
  \( D_{S_i,S_j}^{t} \) & Data amount transferred from \( S_i \) to \( S_j \) at time \(t\) \\
  \( B_{S_i,S_j}^{t} \) & The bandwidth between \( S_i \) and \( S_j \) at time \(t\) \\
  \( B_{S_i,G_j}^{t} \) & The bandwidth between \( S_i \) and \( G_j \) at time \(t\) \\
  \( B_{G_j}^{t} \)& The bandwidth from \( G_j \) to Data Center at time \(t\) \\
  \( Q_{S_i}^{t} \) & The data stored at satellite at time \(t\) \\
  \( Q_{S_i}^{'t} \) & The actual observed queue length for \( S_i \) at time \(t\)\\
  \( P_{S_i}^{t} \) & \(S_i\) generated data between time \(t-1\) and time \(t\) \\
   \( P_{total} \) & \(S_i\) total data volume generated by all satellites\\
  \( T \) & The total number of time slots satellite generated data\\
  
  \hline
  \end{tabular}
  \label{tab:terms_definition}
\end{table}


First, we provide the problem formulation and optimization objective of the flow planning module. Formally, denote the set of satellites as 
\[ \mathcal{S} = \{S_1, S_2, \dots, S_n\} \]

In this paper, we adopt a discrete-time framework, where the interval from time point $t$ to $t+1$ is considered as time unit. Within this time unit, each satellite $S_i$ captures and generates new data $P_{S_i}^{t}$. 
Satellites can transmit data to other satellites through inter-satellite links (ISL) over a interval from time $t$ to time $t+1$. Let ( $B_{S_i,S_j}^{t}$ ) denote the bandwidth available between satellites $S_i$ and $S_j$ during this interval.

Next, the set of ground stations is 
\[ \mathcal{G} = \{G_1, G_2, \dots, G_m\} \]


If the satellite is connected to the ground station within the time interval from $t$ to $t+1$, it can transfer a data amount of $B_{S_i,G_j}^{t}$; if not connected, $B_{S_i,G_j}^{t} = 0$. The satellite is capable of storing data generated by itself and received from other satellites. \(Q_{S_i}^{t}\) represents the amount of data stored on satellite \(S_i\) at time \(t\). Since typical satellites have storage of multiple TBs\cite{tao2023transmitting}, we consider the storage capacity of satellites to be sufficient for data storage.

Our goal is to compute a \textbf{data transmission plan}, which can be formulated as two matrices \(D_{S_i,S_j}^{t}\) and \(D_{S_i,G_k}^{t}\), representing the amount of data satellite \(S_i\) transferred to its neighboring satellite \(S_j\) or to the ground station \(G_k\) within the time interval from \(t\) to \(t+1\). \(D\) is subject to the following constraints:


\begin{itemize}
  \item A satellite communicates with at most one ground station at a time: \\ \(\forall t, S_i, G_j \neq G_k, D_{S_i,G_j}^{t} = 0 \vee D_{S_i,G_k}^{t} = 0\).
  \item A ground station communicates with at most one satellite at a time: \\ \(\forall t,G_k, S_i \neq S_j, D_{S_i,G_k}^{t} = 0 \vee D_{S_j,G_k}^{t} = 0\) 
  \item A satellite’s transfer speed to the ground station cannot exceed GSL bandwidth: \\ \(\forall t, S_i, G_j, D_{S_i,G_j}^{t} \leq B_{S_i,G_j}^{t}\).
  \item A satellite’s transfer speed to another satellite cannot exceed ISL bandwidth:\\  \(\forall t, S_i, S_j, D_{S_i,S_j}^{t} \leq B_{S_i,S_j}^{t}\) where \(i \neq j\).
  \item A satellite cannot transmit more data than it store: \\ \(\forall t, i,(\sum_{j=0}^{n}D_{S_i,S_j}^t+\sum_{k=0}^{m}D_{S_i,G_k}^t) \leq Q_{S_i}^{t} \).
\end{itemize}


The data stored on the satellite can be calculated based on the previously generated and transmitted data as follows:

\begin{equation}
  Q_{S_i}^{t} = \sum_{j=0}^{t-1}P_{S_i}^{j}+\sum_{k=0}^{t-1}\sum_{j=0}^{n}D_{S_i,S_j}^{k} - \sum_{k=0}^{t-1}(\sum_{j=0}^{n}D_{S_i,S_j}^{k}+\sum_{j=0}^{m}D_{S_i,G_j}^{k})
  \end{equation}
  


\textbf{Optimization Objective:} Our optimization goal is to minimize the time taken to download all the data from the satellites after the last data generation time slot $T$. Mathematically, we define the download completion time $T_{\text{Download}}$ as the shortest time such that the total stored data on all satellites at time $T_{\text{Download}}$ is 0. Thus, we can express our optimization problem as:


\begin{equation}
T_{\text{Download}} = \min \{t \in \mathbb{N} \mid t \geq T \text{ and } \sum_{i=1}^{n} Q_{S_i}^{t} = 0 \}
\end{equation}

\subsection{Transmission Plan Generating}
\begin{figure}[h]
  \centering
  \begin{subfigure}[b]{0.43\textwidth} 
  \includegraphics[width=\linewidth]{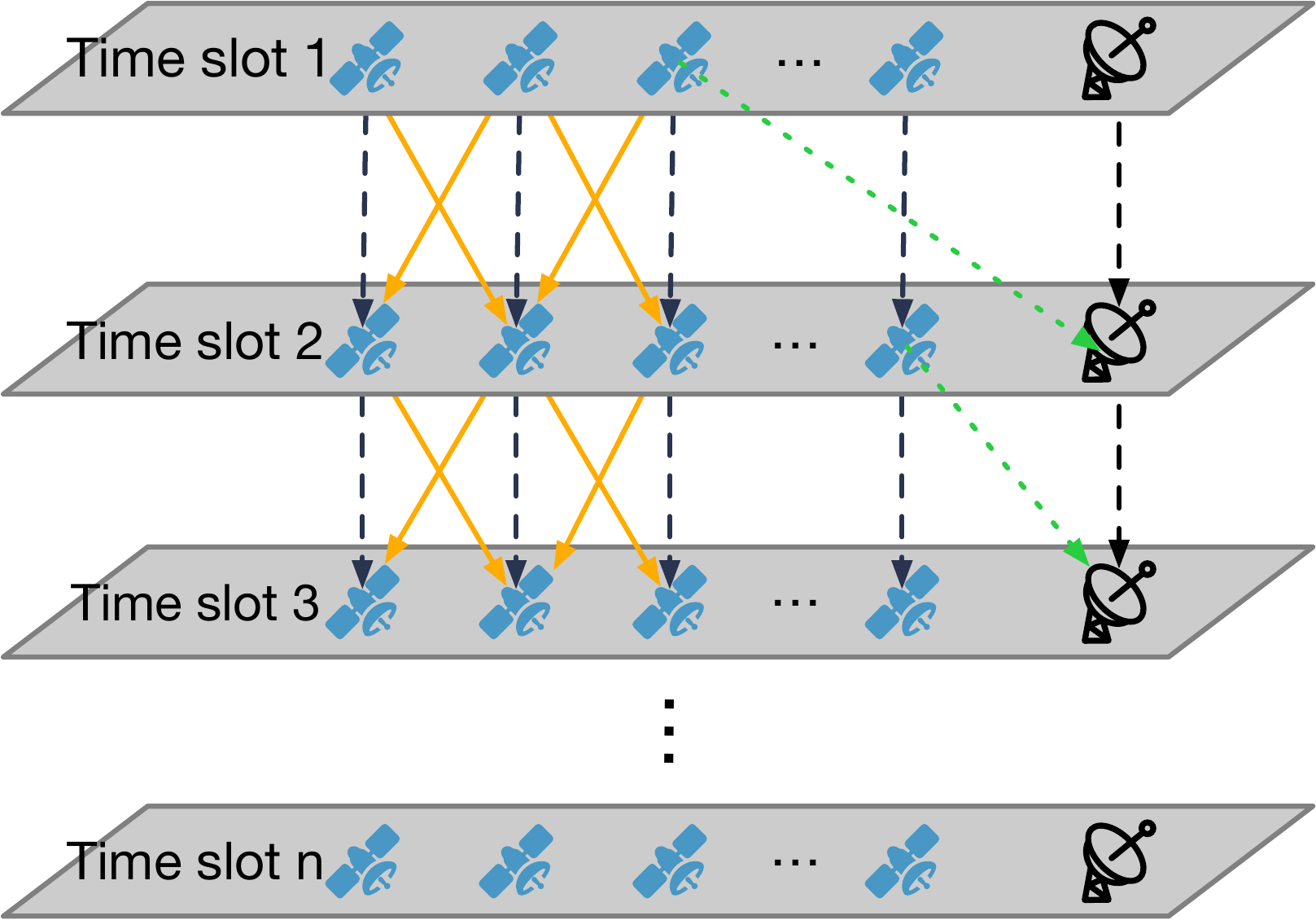}
  \caption{Layered network flow graph with edges between satellite node and ground station node}
  \label{fig:timeslotgraph}

\end{subfigure}
\hfill 
\begin{subfigure}[b]{0.55\textwidth} 
  \includegraphics[width=\linewidth]{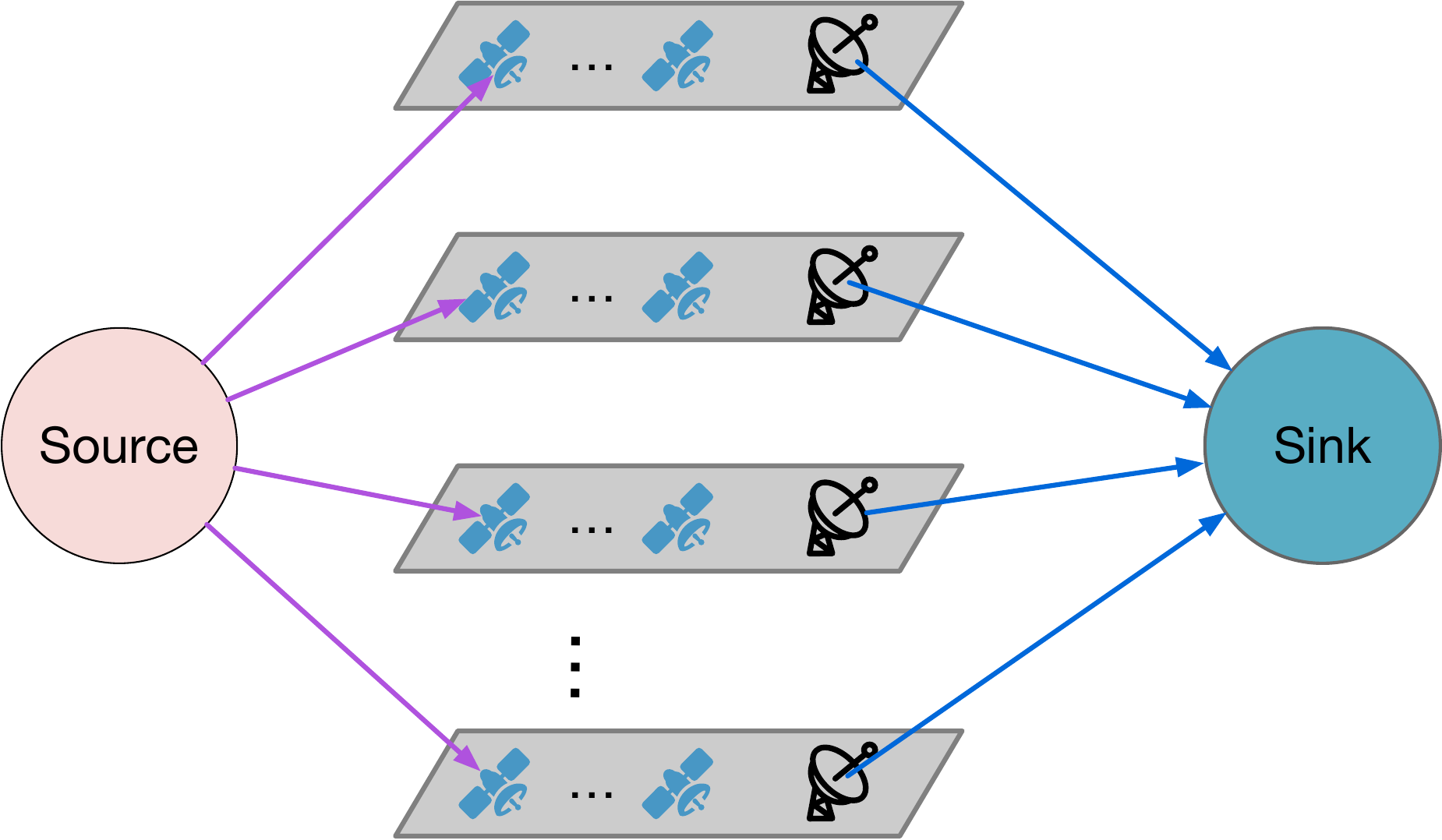}
  \caption{Layered network flow graph with source and sink node}
  \label{fig:timeslotgraphss}
\end{subfigure}
\caption{Layered network flow graph}

\end{figure}

Generating the transmission plan requires solving the layered network flow graph using the Min-Cost Max-Flow algorithm. First, as illustrated in Fig.~\ref{fig:timeslotgraph}, we treat each time slot required for the transmission plan as a layer of network flow graph. For each satellite $S_i$ and ground station $G_j$, we create corresponding nodes $S_i^t$ and $G_j^t$ in each graph layer.



Firstly, for data generation, if satellite \( S_i \) produces data between \( T_{j-1} \) and \( T_j \), an edge from the source node to \( S_i^{t} \) is established with a flow capacity of \( P_{S_i}^{t} \), representing the generated data.
Simultaneously, for each ground station \( G_i \) at every time slot, an edge is drawn from \( G_i^{t} \) to the sink node with a capacity of \( B_{G_i}^t \), depicting the bandwidth available for uploading data to the cloud for processing.

\begin{algorithm}[!ht]
\caption{Adaptive Doubling Strategy for Transmission Plan}
\label{algo:adaptive}
\begin{algorithmic}[1]
\State $T_{gen} \gets 0, M \gets 1, G \gets \emptyset$
\While{True}
    \State $G \gets \text{Generate layered flow graph (length=} T_{gen}+M\text{)}$
    \State $P_{\text{transmit}} \gets \text{Maxflow}(G, \text{source, sink})$
    \If{$P_{\text{transmit}} == P_{\text{total}}$}
        \State $T_{gen} \gets T_{gen} + \lfloor M/2 \rfloor$
        \State \textbf{break}
    \Else
        \State $M \gets M*2$
    \EndIf
\EndWhile
\State \Return $T_{gen}+1$
\end{algorithmic}
\end{algorithm}

We use the following step to create the edge between the satellite nodes and the ground station node:
\begin{enumerate}

    \item Edges between satellite \( S_j^t \) and ground station \( G_i^{t+1} \) nodes are drawn with a flow capacity \( B_{S_j,G_i}^{t} \) if they are scheduled to communicate within the time slot \(t\) to \(t+1\).
    \item If satellite \(S_i\) can transmit data to \(S_j\) within the time slot \(t\) to \(t+1\), an edge from \(S_i^t\) to \( S_j^{t+1}\) is created with a capacity \( B_{S_i,S_j}^{t} \).
    \item Edges of infinite capacity link the same satellite and ground station nodes across consecutive time slots \( S_i^{t} \) to \( S_i^{t+1} \) and \( G_i^{t} \) to \( G_i^{t+1} \), symbolizing the data store capability.
\end{enumerate}

For each edge in the network flow graph, we assign a cost of $1$ to encourage rapid data transfer to the ground.


\subsection{Adaptive Transmission Plan Update}
For packet loss and link failures, we use the Satellite Queue Deviation Index (SQDI) to assess the need to update the transmission plan. The SQDI is calculated as
\begin{equation}
\text{SQDI} =  \frac{1}{n} \sum_{i=1}^{n} \left| {Q'}_{i}^{t} - Q_{i}^{t} \right| + \max_{i=1}^{n} \left| {Q'}_{i}^{t} - Q_{i}^{t} \right|
\end{equation}
which triggers a plan update when it exceeds the threshold \(\theta\). To update the transmission plan, we need to reconstruct the network flow graph from the current time slot, adding edges from the source to each satellite node \(S_{i}^{t}\) at the beginning of the graph. These edges have a capacity of \( {Q'}_{i}^{t} \), representing the current data status of each satellite.
\subsection{Generation Acceleration}
To address the challenge of increased latency from extended computation times, we have developed an adaptive doubling strategy in Algorithm~\ref{algo:adaptive}. This strategy efficiently identifies the optimal length \(T_{gen}\) for transmission plan generation, where \(T_{gen}\) is the minimum time necessary for completing data downloading.





\section{EXPERIMENTAL EVALUATION}

\begin{figure}[!t]
  \centering
    \includegraphics[width=\textwidth]{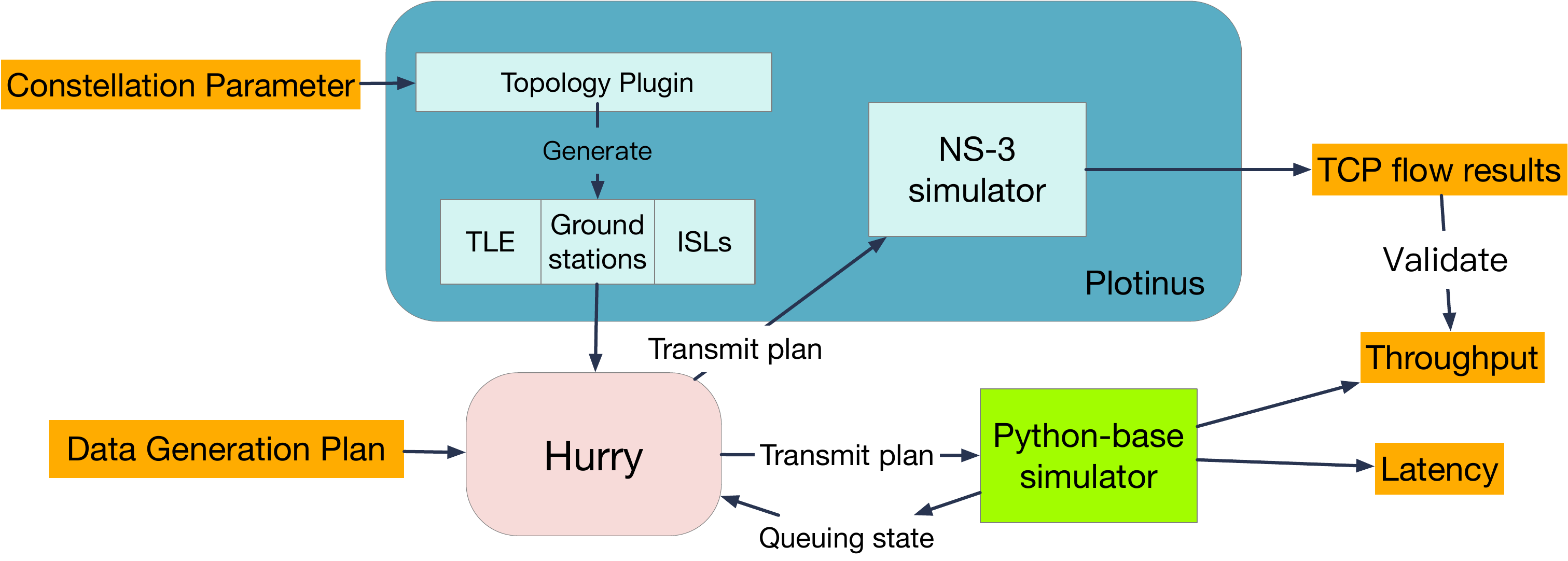}
    \caption{The evaluation process of Hurry}
    \label{fig:evaluates}
\end{figure}

\subsection{Evaluation Environment}
We implement the Hurry framework as a plugin into the Plotinus~\cite{Gao2024Plotinus} to validate the performance of our algorithm. The specific validation process is illustrated in Fig.~\ref{fig:evaluates}, where we use Constellation Parameters as input to the Topology Plugin in Plotinus, which generates the satellite TLE, ground station locations, and inter-satellite link topology information required by Hurry. We use starlink~\textit{shell1} as our experimental satellite constellation.
We set the bandwidth for inter-satellite and satellite-to-ground links for light load scenarios at 10 Gbps. In moderate load scenarios, the bandwidth is set to 5 Gbps for both types of links, and for heavy load scenarios, it is reduced to 1 Gbps.

\subsection{Baseline Algorithms}
We use the following algorithms as baselines to compare the performance of our algorithm under different scenarios: 1) \textbf{CoDld}: The original CoDld algorithm; 2) \textbf{CoDld Modify}: An adjustment to the original CoDld algorithm that incorporates iterative logic to enable data offloading and downloading cycles as soon as satellites generate data; 3) \textbf{Greedy with ISL}: Satellites use Floyd algorithm to find the shortest path (in terms of hops) to the ground station; 4) \textbf{Greedy without ISL}: This algorithm only initiates data transmission when a satellite has a direct connection to the ground station.

  \subsection{Overall Performance}
\begin{table}[!t]
\begin{tabular}{|c|ccccccccc|}
\hline
                   & \multicolumn{9}{c|}{Downloading time (s)}                                                                                                                                                                                                        \\ \hline
GSL bandwidth      & \multicolumn{3}{c|}{1Gbps}                                                            & \multicolumn{3}{c|}{5Gbps}                                                            & \multicolumn{3}{c|}{10Gbps}                                      \\ \hline
ISL bandwidth      & \multicolumn{1}{c|}{1Gbps} & \multicolumn{1}{c|}{5Gbps} & \multicolumn{1}{c|}{10Gbps} & \multicolumn{1}{c|}{1Gbps} & \multicolumn{1}{c|}{5Gbps} & \multicolumn{1}{c|}{10Gbps} & \multicolumn{1}{c|}{1Gbps} & \multicolumn{1}{c|}{5Gbps} & 10Gbps \\ \hline
Greedy (no ISL) & \multicolumn{1}{c|}{30008} & \multicolumn{1}{c|}{30008} & \multicolumn{1}{c|}{30008}  & \multicolumn{1}{c|}{29735} & \multicolumn{1}{c|}{29735} & \multicolumn{1}{c|}{29735}  & \multicolumn{1}{c|}{29735} & \multicolumn{1}{c|}{29735} & 29735  \\ \hline
Greedy with ISL    & \multicolumn{1}{c|}{18029} & \multicolumn{1}{c|}{15854} & \multicolumn{1}{c|}{15853}  & \multicolumn{1}{c|}{18029} & \multicolumn{1}{c|}{15618} & \multicolumn{1}{c|}{15423}  & \multicolumn{1}{c|}{18029} & \multicolumn{1}{c|}{15618} & 15369  \\ \hline
CoDld              & \multicolumn{1}{c|}{6594}  & \multicolumn{1}{c|}{6021}  & \multicolumn{1}{c|}{6021}   & \multicolumn{1}{c|}{7037}  & \multicolumn{1}{c|}{6660}  & \multicolumn{1}{c|}{6660}   & \multicolumn{1}{c|}{7184}  & \multicolumn{1}{c|}{6850}  & 6850   \\ \hline
CoDld Modify       & \multicolumn{1}{c|}{6456}  & \multicolumn{1}{c|}{5727}  & \multicolumn{1}{c|}{5727}   & \multicolumn{1}{c|}{7043}  & \multicolumn{1}{c|}{6594}  & \multicolumn{1}{c|}{6594}   & \multicolumn{1}{c|}{6989}  & \multicolumn{1}{c|}{5943}  & 5943   \\ \hline
Hurry              & \multicolumn{1}{c|}{65}    & \multicolumn{1}{c|}{65}    & \multicolumn{1}{c|}{65}     & \multicolumn{1}{c|}{52}    & \multicolumn{1}{c|}{26}    & \multicolumn{1}{c|}{26}     & \multicolumn{1}{c|}{52}    & \multicolumn{1}{c|}{25}    & 25     \\ \hline
\end{tabular}
   \captionsetup{skip=10pt}
    \caption{Algorithm performance for satellite data downloads. \textnormal{Each entry in the table represents the time it takes for the corresponding algorithm to download 99\% of 0.25GB of data generated by each satellite over a 10-second period.}}
    \vspace{-1cm}
    \label{tab:satellite_data_download}
\end{table}

To verify the performance of the proposed algorithm, we test the time required to download satellite data under different link bandwidth combination scenarios. The experimental results are shown in Table \ref{tab:satellite_data_download}. Our algorithm consistently outperforms the baseline algorithm in various scenarios, reducing the download time by more than 99\%. Our algorithm's download time under heavy load exceeds twice that of medium and light loads. However, in the CoDld algorithm and its modified version, the download times across different bandwidths essentially remain consistent. For the Greedy with ISL algorithm, its download time is approximately three times that of the CoDld algorithm. When the ISL bandwidth is 5 Gbps or higher, its download time can be reduced by 17\% compared to when the bandwidth is 1 Gbps.

  \begin{figure}[h] 
    \centering
    \includegraphics[width=\linewidth]{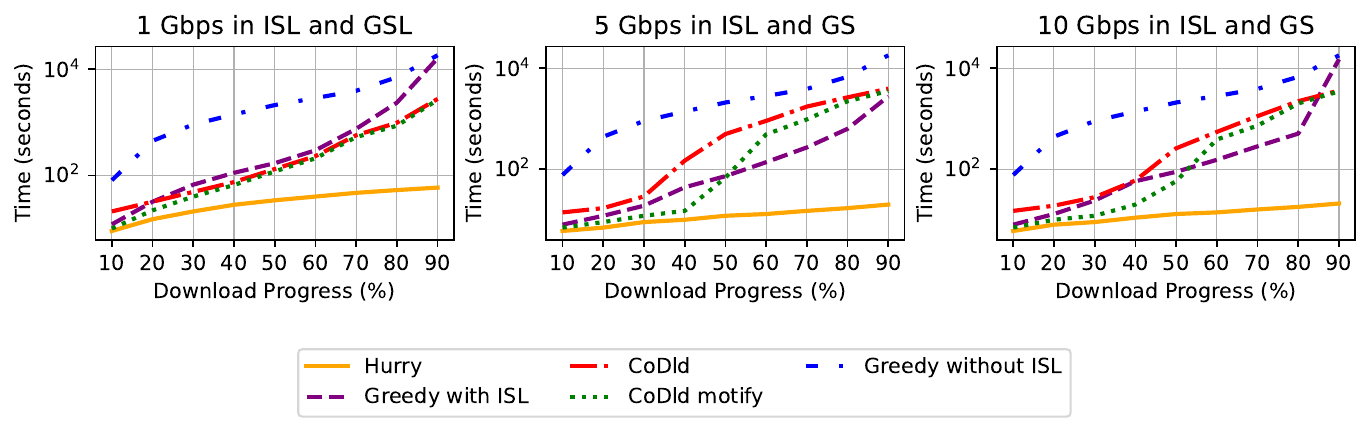}
    \caption{Time to reach download progress for different inter-satellite link and ground station-satellite link bandwidths.}
    \vspace{-0.4cm}
    \label{fig:progress_time}
\end{figure}
  
  In Fig.\ref{fig:progress_time}, we present the time required by the Hurry and baseline algorithms to reach corresponding progress milestones in the download process under various combinations of ISL and GSL bandwidths. Hurry maintains an almost constant rate from the beginning to the end of the downloading, whereas the time taken by other algorithms to complete corresponding progress milestones sharply increases as the download proceeds. At 10\% transmission progress, the Hurry, CoDld Modify, and Greedy with ISL algorithms require roughly the same amount of time. As the transmission continues, the gap between the CoDld Modify and the original CoDld algorithms gradually narrows in the graph, with the time taken for corresponding progress levels showing exponential growth. On the other hand, the time for corresponding progress in the Greedy with ISL algorithm initially grows exponentially but then completes the transmission rapidly towards the end. The Greedy without ISL algorithm takes significantly longer for data transmission than all other algorithms. For the flow planning algorithm, the primary bottleneck under heavy load is the bandwidth of the GSL, while under light and moderate loads, most of the time is consumed by the delay caused by data transmission between satellites.

\subsection{Performance of Long-time Generation}

\begin{figure}[t] 
  \centering
  \includegraphics[width=\linewidth]{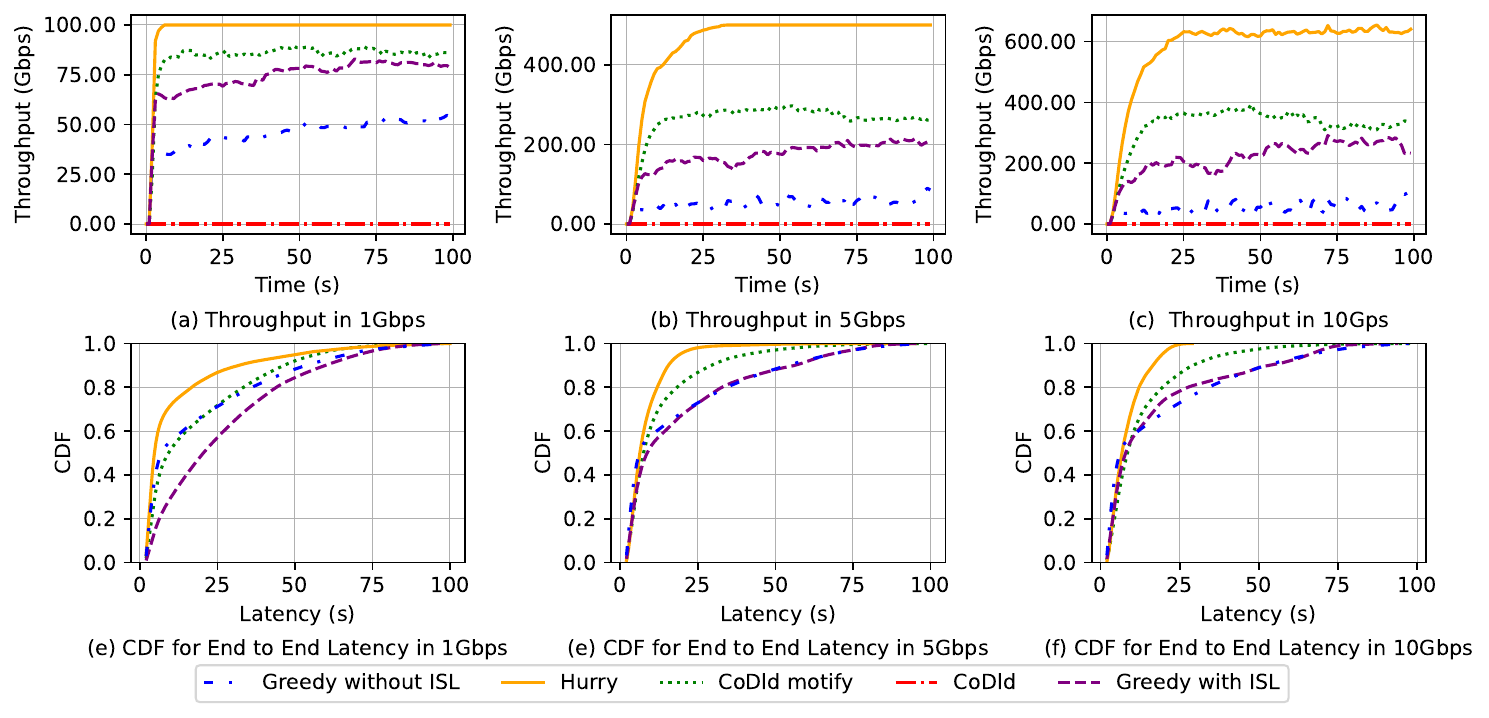}
  
  \caption{Throughput over time in different ISL and GSL bandwidths, and latency CDF in different scenarios.}
  \vspace{-0.4cm}
  \label{fig:throughput_long}
\end{figure}

We test the performance of our algorithm and the baseline algorithms in a scenario where each satellite continuously generates data at a rate of 400Mbps. As shown in Fig.~\ref{fig:throughput_long}(a)(b)(c), the throughput indicates that our algorithm, under a combination of 1Gbps ISL and GSL bandwidths, improved the overall throughput by approximately 11\% compared to the CoDld Modify algorithm, which has the highest throughput among the baselines. At bandwidths of 5Gbps and 10Gbps, the throughput increased by 66\% and 50\%, respectively.

During transmission, our algorithm and CoDld Modify maintain consistent throughput after reaching their maximum values. The throughput of the Greedy with ISL algorithm gradually increases over time. The throughput of the Greedy without ISL algorithm slowly increases at low GSL bandwidths and remains unchanged at medium and high GSL bandwidths. The throughput of the original CoDld algorithm is consistently zero due to its lack of support for dynamic data generation. The CoDld Modify algorithm, affected by the Proximal Station Bias Degradation (PSBD) phenomenon, can only effectively transmit data within a few hops of the ground station.

Fig.~\ref{fig:throughput_long}(d)(e)(f) illustrates the CDF graph of the packets transmitted by the flow planning algorithm and the baseline in the continuous data generation experiment. Our algorithm exhibits lower latency than the baseline in every bandwidth experiment, with the delay further decreasing as the bandwidth of ISL and GSL increases. The latency of the CoDld modify algorithm is superior to that of Greedy with ISL across all three bandwidths due to the fact that Greedy with ISL encounters more congestion phenomena during data transmission.

\section{CONCLUSION}

In this paper, we introduce the Hurry framework and the flow planning algorithm, which significantly reduce the satellite data downloading time in Low-orbit Mega-Constellation Networks (LMCN), identifying and thoroughly analyzing the Proximal Station Bias Degradation (PSBD) phenomenon present in the existing CoDld algorithm within LMCN. By modeling and mapping the changes in satellite topology and data transmission onto Time-Expanded Graphs, the flow planning algorithm is able to generate an efficient transmission plan. In our experiments conducted using the Plotinus satellite digital twin system, in the fixed data volume download evaluation, Hurry achieves 100\% completion of the download task while the CoDld has reached only 44\% download progress. Furthermore, in scenarios of continuous data generation, Hurry increases throughput by 11\% to 66\%, highlighting its capability to adjust transmission plan for optimal efficiency dynamically. 
As a potential future direction, we are looking forward to extending our Hurry to improve the performance of various applications such as distributed learning systems~\cite{lin2024efficient,lin2024adaptsfl,lin2024split}, large language models~\cite{fang2024automated,lin2023pushing}, ISAC~\cite{hu2023holofed}, etc in LEO satellite networks.

\bibliographystyle{splncs04}

\bibliography{main}

\end{document}